\documentclass[twocolumn,showpacs,preprintnumbers,amsmath,amssymb,footinbib,APS]{revtex4}

\usepackage{dcolumn}
\usepackage{bm}
\usepackage[latin1]{inputenc}
\usepackage[spanish,english]{babel}
\usepackage{amsfonts}
\usepackage{amssymb}
\usepackage{graphicx}

\newcommand{\R}{\mathcal{R}}

\begin{document}

\title{ {\bf Re-examination of Polytropic Spheres in Palatini $f(R)$ Gravity}}
\author{Gonzalo J. Olmo\thanks{olmo@iem.cfmac.csic.es}}
\affiliation{ {\footnotesize Instituto de Estructura de la Materia, CSIC, Serrano 121, 28006 Madrid, Spain }\\ and \\ {\footnotesize Departamento de Física Teórica and
IFIC, Centro Mixto Universidad de Valencia-CSIC.
    Facultad de Física, Universidad de Valencia,
        Burjassot-46100, Valencia, Spain. }}


\date{October 20th, 2008}

\begin{abstract}
We investigate spherically symmetric, static matter configurations with polytropic equation of state for a class of $f(R)$ models in Palatini formalism and show that the surface singularities recently reported in the literature are not physical in the case of Planck scale modified lagrangians.  In such cases, they are just an artifact of the idealized equation of state used. In fact, we show that for the models $f(R)=R\pm\lambda R^2$, with $\lambda$ on the order of the Planck length squared, the presence of a single electron in the Universe would be enough to cure all stellar singularities of this type. From our analysis it also follows that the stellar structure derived from these lagrangians is virtually undistinguishable from that corresponding to General Relativity. For ultraviolet corrected models far from the Planck scale, however, the surface singularities may indeed arise in the region of validity of the polytropic equation of state. This fact can be used to place constraints on the parameters of particular models. 
\end{abstract}


\maketitle

\section{Introduction}

The study of modified theories of gravity is interesting for many reasons. General Relativity (GR) has stood as a monument to the genius of Albert Einstein for almost a century. However, despite the astonishing observational and experimental success of this theory \cite{WIL-liv05}, there exist many theoretical aspects which make us consider the possibility of departures from it in certain regimes. In fact, one would like to remove physically undesired aspects such as the Big Bang or black hole singularities. These are aspects which clearly point towards high energy modifications of the theory motivated by quantum gravity effects. On the other hand, the recently observed cosmic speedup \cite{Tonry-Knop} suggests that something is needed in the Friedman-Robertson-Walker equations to account for the apparent positive acceleration of the expansion rate. This requires a modification of the equations at low energies. It could come in the form of a new energy source dominant over the other matter sources at very low densities \cite{Dark-energy}, or as a modification of the field equations which would be suppressed above a certain cosmic curvature scale \cite{mod-grav-basic}. It is this latter possibility which has impulsed the study of modified theories of gravity of the $f(R)$ type in the recent literature \cite{mod-grav-long} (see also \cite{f_reviews}). \\

When the Einstein-Hilbert lagrangian is promoted to a non-linear function of the scalar curvature one is dealing with an $f(R)$ theory of gravity. When this is done, one must specify whether the connection is still defined in terms of the metric as the Levi-Civita connection (metric formalism) or if it is an independent geometrical field (Palatini formalism). One should note that, on pure geometrical grounds, metric and connection are completely independent and equally fundamental objects. However, the literature has traditionally given a dominant role to the metric over the connection. In this paper we will focus on $f(R)$ theories in Palatini formalism. In these theories, the field equations show that the independent connection is not a dynamical field in the sense that it does not satisfy an independent second-order partial differential equation. Instead, its field equation can be solved in terms of an auxiliary metric which is conformally related with the usual (or physical or Jordan frame) metric. As we will see, the conformal factor is a local function of the trace, $T$, of the energy-momentum tensor $T_{\mu\nu}$ of the matter. Thus, the independent connection, despite being non-dynamical in the usual sense, does mediate new interactions between the matter and geometry. The resulting equations for the metric can be written in Einstein form with a modified right hand side in which the trace $T$ plays a role non-existing in GR. We will discuss later in depth the implications of such terms in the dynamics of these theories. \\

The goal of this paper is to contribute and shed some new light on a problem discussed in recent literature that seems to affect seriously the theoretical viability of {\it all} $f(R)$ models in Palatini formalism. The problem consists on the existence of curvature singularities near the surface of spherically symmetric, static matter configurations with polytropic equation of state (EoS) and polytropic index within the range $3/2<\gamma<2$. Its existence and possible consequences were first reported by Barausse et al. in \cite{Sot08a} using previous results of Kainulainen et al. \cite{Kainu07a}. The problem was reconsidered by Kainulainen et al. in \cite{Kainu07b}, where it was argued that the singularities had more to do with the peculiarities of the EoS used than with the theory of Palatini $f(R)$ gravity. They found that for neutron stars the tidal acceleration due to the singularity becomes equal to the Schwarzschild value of GR only at a distance  $\sim 0.3$ fermi from the surface of the star, which makes unrealistic the use of a polytropic EoS. However, Barausse et al. replied in \cite{Sot08b} that the fluid approximation is still valid on the scale at which the tidal forces diverge just below the surface of a polytropic sphere in the case of the generic functions $f(R)$ considered.\\We reconsider the problem and independently re-derive the structure equations for spherically symmetric, static systems. We study the physically admissible models $f(R)=R\pm \lambda R^2$, with $\lambda$ a constant of order the Planck length squared, and estimate the density scale $\rho_s$ at which the potentially divergent terms become of order unity. We find that, for our choice of $\lambda$, $\rho_s$ is well below any physically accessible density scale. In this case, one can conclude that the mathematical singularity found in \cite{Sot08a} and \cite{Sot08b} is just an artifact due to the use of the polytropic EoS beyond its natural regime of validity. For other choices of $\lambda$, however, the singularity can effectively arise at densities sufficiently high as to admit the validity of the polytropic EoS. Our approach and results can thus be used to place bounds on the allowed range of values of $\lambda$ and to constrain other $f(R)$ models. \\

The paper is organized as follows. In section II we present the action, derive the field equations, and discuss their physical properties and new aspects as compared to GR. Section III presents the structure equations for a spherically symmetric, static matter configuration for a generic, unspecified lagrangian $f(R)$. In section IV we study the behavior of the metric near the surface for systems with polytropic EoS. We conclude with a summary and discussion. 

\section{Action and Field Equations}

Let us begin by defining the action of Palatini theories
\begin{equation}\label{eq:Pal-Action}
S[{g},\Gamma ,\psi_m]=\frac{1}{2\kappa^2}\int d^4
x\sqrt{-{g}}f({R})+S_m[{g}_{\mu \nu},\psi_m]
\end{equation}
Here $f({R})$ is a function of ${R}\equiv{g}^{\mu \nu }R_{\mu \nu }(\Gamma )$, with $R_{\mu \nu }(\Gamma )$ given by
$R_{\mu\nu}(\Gamma )=-\partial_{\mu}
\Gamma^{\lambda}_{\lambda\nu}+\partial_{\lambda}
\Gamma^{\lambda}_{\mu\nu}+\Gamma^{\lambda}_{\mu\rho}\Gamma^{\rho}_{\nu\lambda}-\Gamma^{\lambda}_{\nu\rho}\Gamma^{\rho}_{\mu\lambda}$
where $\Gamma^\lambda _{\mu \nu }$ is the connection. The matter action $S_m$ depends on the matter fields $\psi_m$, the metric $g_{\mu\nu}$, which defines the line element $ds^2=g_{\mu\nu}dx^\mu dx^\nu$, and its first derivatives (Christoffel symbols). The matter action does not depend on the connection $\Gamma^\lambda _{\mu \nu }$, which is seen as an independent field appearing only in the gravitational action (this condition is not essential and can be relaxed at the cost of introducing a non-vanishing torsion). Varying (\ref{eq:Pal-Action}) with respect to the metric $g_{\mu\nu}$ we obtain
\begin{equation}\label{eq:met-var-P}
f'(R)R_{\mu\nu}(\Gamma)-\frac{1}{2}f(R)g_{\mu\nu}=\kappa ^2T_{\mu
\nu }
\end{equation}
where $f'(R)\equiv df/dR$. From this equation we see that the scalar $R$ can be solved as an algebraic function of the trace $T$. This follows from the trace of
(\ref{eq:met-var-P})
\begin{equation}\label{eq:trace-P}
f'(R)R-2f(R)=\kappa ^2T,
\end{equation}
The solution to this algebraic equation will be denoted by $R=\R(T)$. The variation of (\ref{eq:Pal-Action}) with respect to $\Gamma^\lambda _{\mu \nu }$ must vanish independently of (\ref{eq:met-var-P}) and gives
\begin{equation}\label{eq:con-var}
\nabla_\rho^\Gamma  \left[\sqrt{-g}\left(\delta ^\rho _\lambda
f'g^{\mu \nu }-\frac{1}{2}\delta ^\mu _\lambda f'g^{\rho
\nu }-\frac{1}{2}\delta^\nu_\lambda f'g^{\mu
\rho}\right)\right]=0
\end{equation}
where $\nabla_\rho^\Gamma$ is the derivative operator associated to $\Gamma$ and $f'\equiv f'(\R[T])$ is a function of the matter fields. This equation leads to 
\begin{equation}\label{eq:Gamma-1}
\Gamma^\lambda_{\mu \nu }=\frac{t^{\lambda \rho
}}{2}\left(\partial_\mu t_{\rho \nu }+\partial_\nu
t_{\rho \mu }-\partial_\rho t_{\mu \nu }\right)
\end{equation}
where  $t_{\mu \nu }\equiv f' g_{\mu \nu }$, and $f'\equiv \frac{df}{dR}$. Inserting this solution for $\Gamma$ into (\ref{eq:met-var-P}) we find the following modified Einstein equations 
\begin{eqnarray}\label{eq:neweinstein}
G_{\mu \nu}(g) &=& \frac{\kappa^2}{f'} T_{\mu \nu} - \frac{\R f' - f}{2 f'} g_{\mu \nu} + \frac{1}{f'}\left(\nabla_\mu \nabla_\nu f' - g_{\mu \nu} \Box f'\right) \nonumber\\ 
&& - \frac{3}{2 f'^2} \left(\partial_\mu f' \partial_\nu f' - \frac{1}{2} g_{\mu \nu} (\partial f')^2\right) 
\end{eqnarray}
where now $\nabla_\mu$ is the usual covariant derivative of $g_{\mu\nu}$. \\

\subsection{Physical aspects of the field equations}

First of all, it is important to note that the right hand side of (\ref{eq:neweinstein})
behaves like a modified energy-momentum tensor in which the trace $T$ plays a role non-existing in GR via the terms $\R,f(\R)$ and the various derivatives of $f(\R)$. Thus, though the equations for the metric are still second order, the dynamics in regions with pronounced gradients in the matter fields ($\partial f'$ and $\nabla_\mu\nabla_\nu f'$) might considerably depart from that of standard GR. In contrast, the vacuum solutions of these equations are exactly the same as in vacuum GR with a cosmological constant.\\

To better understand the role of the gradients of $f'$ in (\ref{eq:neweinstein}),  it is useful to re-express the field equations in terms of the auxiliary metric $t_{\mu\nu}$ that defines the connection $\Gamma$ in (\ref{eq:Gamma-1}). In terms of this metric, the equations become
\begin{equation}\label{eq:G-tmn}
G_{\mu\nu}(t)=\frac{\kappa^2}{f'}T_{\mu\nu}-\frac{[\R f'-f]}{2(f')^2}t_{\mu\nu}
\end{equation}
The advantage of using $t_{\mu\nu}$ is that the gradients $\partial_\mu f'$ and $\nabla_\mu\nabla_\nu f'$ present in (\ref{eq:neweinstein}) are completely absent in (\ref{eq:G-tmn}), which allows for a better comparison with GR. The structure of these equations resembles that of GR plus a trace-dependent term $\frac{[\R f'-f]}{2(f')^2}$, which becomes a cosmological constant in vacuum. The solution for $t_{\mu\nu}$ will thus involve integrals over the matter sources which, like in GR, represent the fact that larger amounts of matter contribute more to the curvature of space-time. What is then the role of the gradients $\partial_\mu f'$ and $\nabla_\mu\nabla_\nu f'$ of (\ref{eq:neweinstein})? The answer is simple: since the physical metric is $g_{\mu\nu}=\frac{t_{\mu\nu}}{f'}$, the gradients of $f'$ in (\ref{eq:neweinstein}) introduce a local dependence of the metric on the local matter-energy distribution. The metric is not only determined by the total amount of matter, also the local matter-energy density affects its properties via $f'(\R(T))$. This dependence on the local energy density distribution manifests itself with clarity in $f(R)$ models sensitive to low energy scales. In particular, it has been shown that all such models would have dramatic effects for the stability of matter \cite{Olmo07}. For the model $f(R)=R-\mu^4/R$, the ground state of Hydrogen would desintegrate in less than two hours \cite{Olmo08a}, which rules out that model and all models sensitive to ultra-low matter density scales. The reason for this has to do with the strong gravitational backreaction induced by $f'$ near the zeros of the atomic wave functions, where the characteristic ultra-low density scale $\rho_\mu\equiv\mu^2/\kappa^2\sim 10^{-26} \ g/cm^3$ can be effectively reached. In fact, for atoms (and many other not so small systems), the total amount of matter is very small and leads to $t_{\mu\nu}\approx \eta_{\mu\nu}$. However, if $f'$ is sensitive to ultra-low densities such as $\rho_\mu$, the physical metric $g_{\mu\nu}=\eta_{\mu\nu}/f'$ departs from $\eta_{\mu\nu}$ when the matter density $m_e|\psi(x)|^2$ drops below $\rho_\mu$.  This is at the heart of the matter instabilities found in \cite{Olmo07,Olmo08a}.\\

On the other hand, the dependence of $g_{\mu\nu}$ on the local matter-energy distribution also imposes constraints on the sources. Since the Riemann tensor involves second-order derivatives of the metric, it follows that a smooth geometry must be one with continuous and differentiable $f'$ up to second order, $\partial^2 g_{\mu\nu} \sim (\partial^2 f') t_{\mu\nu} \sim (\partial f')^2 t_{\mu\nu}$, which implies that $T$ must be differentiable up to that order at least \footnote{In a Universe like ours, described by Quantum Mechanics at microscopic scales and in which objects with sharp edges are classical idealizations, this condition on the sources should not be seen as a drawback of the theory.}. An immediate consequence of this restriction is that the naive spherically symmetric constructions of GR with matter profile of the form 
\begin{equation}
\rho=\left\{\begin{array}{lr} \rho_0 & \makebox{ if } r<R_0 \\
                                 0   & \makebox{ if } r\ge R_0 \end{array}\right.
\end{equation}
are, strictly speaking, not valid for a generic Palatini $f(R)$. This idealized construction, used as such in \cite{Kainu07a}, can be easily adapted to Palatini $f(R)$ by conveniently modifying the density profile near the surface to smoothly interpolate between the interior $\rho_0$ and the exterior vacuum. Otherwise, the discontinuous density profile would yield undesired singularities. Nonetheless, intuition tells us that such a refinement should be (from a practical point of view) unnecessary for theories with high curvature corrections, for which the contribution to the metric due to the matter contained in the outermost layers of the star should be negligible and then one could effectively use GR. However, it is precisely the existence of singularities on the surface of stellar objects that motivates this paper. In the next sections we will study in detail the properties and nature of such (highly unexpected and counterintuitive) singularities. 

\section{Stellar Structure equations}

Using the following line element
\begin{equation}
ds^2=-A(r)e^{2\psi(r)}dt^2+\frac{1}{A(r)}dr^2+r^2d\Omega^2
\end{equation}
with $A(r)=1-2M(r)/r$, the combinations $G_t^t\pm G_r^r$ of (\ref{eq:neweinstein}) lead to
\begin{eqnarray} 
\left(\frac{2}{r}+\frac{f'_r}{f'}\right)\psi_r &=& \frac{\kappa^2(P+\rho)}{f'A(r)}+\left[\frac{f'_{rr}}{f'}-\frac{3}{2}\left(\frac{f'_r}{f'}\right)^2\right]\label{eq:psi}\\
\frac{1}{r}\left(\frac{2}{r}+\frac{f'_r}{f'}\right)M_r &=& \frac{\kappa^2(\rho+V/2)}{f'}+ A(r)\left[\frac{f'_{rr}}{f'}-\right.\label{eq:Mr}\\
&-&\left.\frac{3}{4}\left(\frac{f'_r}{f'}\right)^2+\frac{f'_{r}}{f'}\frac{[r-M(r)]}{r[r-2M(r)]}\right] \nonumber 
\end{eqnarray}
where $f'\equiv df/dR$, $f'_r\equiv df'/dr$,  $f'_{rr}\equiv d^2f'/dr^2$,  $\psi_r\equiv d\psi/dr$,  $M_r\equiv dM/dr$, and $V\equiv \R f'-f$.
These two equations, plus the conservation equation 
\begin{equation}\label{eq:P}
\frac{dP}{dr}=-(\rho+P)\left(\psi_r+\frac{A_r}{2A}\right) \ ,
\end{equation}
plus an equation of state $\rho=\rho(P)$ suffice to determine the structure of static, spherically symmetric objects. The equations corresponding to GR are recovered in the limit $f'=1$.\\ 

In order to proceed with the algebraic manipulations and extract information about the new physics hidden in equations (\ref{eq:psi}), (\ref{eq:Mr}), and (\ref{eq:P}), we must substitute (\ref{eq:psi}) and (\ref{eq:Mr}) into (\ref{eq:P}) to get
\begin{eqnarray}\label{eq:P2}
\frac{dP}{dr}&=&-\frac{(\rho+P)}{A(r)}\left[\frac{(\kappa^2P-V/2)}{f'\Delta}+\frac{M(r)}{r^2}-\right.\\ &-&\left.\frac{A(r)}{\Delta}\frac{f'_r}{f'}\left(\frac{3}{4}\frac{f'_r}{f'}+\frac{r-M(r)}{r[r-2M(r)]}\right)\right]\nonumber
\end{eqnarray}
where $\Delta\equiv (2/r+f'_r/f')$. 
The terms in the second line of this equation contain the main contribution due to the modified lagrangian. To deal with them, it is useful to express the matter terms $f'_r$ and $f'_{rr}$ in terms of $P_r$ and $P_{rr}$ explicitly. Since $f'_r=f'' (\partial \R/\partial T)(\partial T/\partial P)P_r$, we can use (\ref{eq:trace-P}) to find 
\begin{equation}\label{eq:dRdT}
\frac{\partial \R}{\partial T}=\frac{\kappa^2}{\R f''-f'} \ .
\end{equation}
This way we can express $f'_r$ as $f'_r=f'_P P_r$, where
\begin{equation}\label{eq:f_P}
f'_P\equiv \frac{\kappa^2 f''}{\R f''-f'}\left(3-\frac{d\rho}{dP}\right)
\end{equation}
and we have used $T=3P-\rho$. Similarly, one finds that $f'_{rr}$ can be expressed as $f'_{rr}=f'_P P_{rr}+f'_{PP}P_r^2$, where
\begin{equation}\label{eq:f_PP}
f'_{PP}=-\frac{\kappa^4 f' f'''}{(\R f''-f')^3}\left(3-\frac{d\rho}{dP}\right)^2-\frac{\kappa^2 f''}{(\R f''-f')}\frac{d^2\rho}{dP^2}
\end{equation}
Fortunately, $f'_{rr}$ does not appear in (\ref{eq:P2}) and the dependence on $f'_r$ leads to a quadratic algebraic equation on $P_r$ which can be solved inmediately. The result can be expressed as follows:
\begin{equation}\label{eq:newP}
\frac{dP}{dr}=-\frac{P^{(0)}_r}{(1-\alpha)}\frac{2}{\left(1+\sqrt{1+\beta P^{(0)}_r}\right)}
\end{equation}
where we have defined
\begin{eqnarray}\label{eq:P0}
P^{(0)}_r &\equiv& \frac{(\rho+P)}{r[r-2M(r)]}\left[M(r)+\left(\kappa^2P-\frac{V}{2}\right)\frac{r^3}{2f'}\right]\\
\alpha &\equiv & \frac{(\rho+P)}{2}\frac{f'_P}{f'} \label{eq:alpha}\\
\beta &\equiv& \frac{3r^2[r-2M(r)]}{2(1-\alpha)^2}\left(\frac{f'_P}{f'}\right)^2(\rho+P) \label{eq:beta}
\end{eqnarray}
Note that $P^{(0)}_r$ coincides, up to the factor $V/2$, with the result corresponding to GR. The terms $\alpha$ and $\beta$ are proportional to $f'_P/f'$ and $(f'_P/f')^2$, respectively, and vanish in the limit of GR, $f'=1$. Using (\ref{eq:newP}) it is now possible to compute $P_{rr}$, which will be needed in (\ref{eq:psi}) and (\ref{eq:Mr}). In terms of $P_r$ and $P_{rr}$, those equations can be cast as
\begin{eqnarray}
\left(1+\frac{r}{2}\frac{f'_P}{f'}P_r\right)\psi_r &=& \frac{\kappa^2(P+\rho)r^2}{2f'[r-2M(r)]}+\frac{r}{2}\left[\left(\frac{f'_{PP}}{f'}-\right.\right. \nonumber\\ &-& \left.\left.\frac{3}{2}\left(\frac{f'_P}{f'}\right)^2\right)P_r^2+\frac{f'_{P}}{f'}P_{rr}\right]\label{eq:psi-P}\\
\left(1+\frac{r}{2}\frac{f'_P}{f'}P_r\right)M_r &=& \frac{\kappa^2(\rho+V/2)r^2}{2f'}+ \frac{r[r-2M(r)]}{2}\times\nonumber \label{eq:Mr-P}\\
&\times& \left[\left(\frac{f'_{PP}}{f'}-\frac{3}{4}\left(\frac{f'_P}{f'}\right)^2\right)P_r^2+\right.\\ &+&\left.\frac{f'_{P}}{f'}P_{rr}+\frac{[r-M(r)]}{r[r-2M(r)]}\frac{f'_{P}}{f'}P_r\right] \nonumber
\end{eqnarray}
We now have all the equations needed to study the structure of spherically symmetric, static configurations in Palatini $f(R)$ theories written in explicit form. 

\section{Polytropes in $f(R)=R\pm \lambda R^2$}

We will now study the properties of equations (\ref{eq:newP}), (\ref{eq:psi-P}), and (\ref{eq:Mr-P}) for a matter distribution with polytropic equation of state
\begin{equation}\label{eq:poly}
\rho(P)=\left(\frac{P}{K}\right)^{1/\gamma}+\frac{P}{\gamma-1}
\end{equation}
in the region close to the surface of such an object, i.e., in the limit $P\to 0$.
Our gravity lagrangian is characterized by the Planck length scale $\lambda=l_P^2$, which is related to a density scale $\rho_\lambda\equiv (\kappa^2\lambda)^{-1}\sim 2\cdot10^{92}$ $g/cm^3$ (we have used $\kappa^2=8\pi G/c^2$). For these models, we find 
\begin{eqnarray}\label{eq:f'}
f'&=& 1\pm2\lambda R=1\mp 2\frac{T}{\rho_\lambda}\\
f'_P&=& \mp\frac{2(3-\rho_P)}{\rho_\lambda} \label{eq:f'p}\\
f'_{PP} &=& \mp\frac{2\rho_{PP}}{\rho_\lambda} \label{eq:f'pp}
\end{eqnarray}
where $T=3P-\rho$ and
\begin{eqnarray}\label{rp_rpp}
\rho_P&=&\frac{1}{\gamma-1}+\frac{1}{\gamma K}\left(\frac{P}{K}\right)^{(1-\gamma)/\gamma} \\
\rho_{PP} &=& \frac{(1-\gamma)}{(\gamma K)^2}\left(\frac{P}{K}\right)^{(1-2\gamma)/\gamma}
\end{eqnarray}
Since we are interested in the regions near the surface, we now study the leading order of the different quantities appearing in (\ref{eq:newP}), (\ref{eq:psi-P}), and (\ref{eq:Mr-P}). For very low pressures,  we find that $\rho\approx \left(\frac{P}{K}\right)^{1/\gamma}$, $\rho_P\approx \frac{1}{\gamma K}\left(\frac{P}{K}\right)^{(1-\gamma)/\gamma}$, and $\rho_{PP}= \frac{(1-\gamma)}{(\gamma K)^2}\left(\frac{P}{K}\right)^{(1-2\gamma)/\gamma}$ remains unchanged. Using this, (\ref{eq:f'}), (\ref{eq:f'p}), and (\ref{eq:f'pp}) become
\begin{eqnarray}
f'&\approx &1\pm 2\frac{\rho}{\rho_\lambda}\approx1\pm\frac{2}{\rho_\lambda}\left(\frac{P}{K}\right)^{1/\gamma}\approx 1 \\
f'_P&\approx& \mp\frac{2}{\rho_\lambda}\frac{1}{\gamma K}\left(\frac{P}{K}\right)^{(1-\gamma)/\gamma} \\
f'_{PP} &=& \mp\frac{2}{\rho_\lambda}\frac{(1-\gamma)}{(\gamma K)^2}\left(\frac{P}{K}\right)^{(1-2\gamma)/\gamma} \label{eq:f'PP_app} \ .
\end{eqnarray}
We also find that (\ref{eq:P0}), (\ref{eq:alpha}), and (\ref{eq:beta}) become
\begin{eqnarray}
P_r^{(0)}& \approx & \frac{M_{tot}}{r(r-2M_{tot})}\left(\frac{P}{K}\right)^{1/\gamma} \label{eq:P0_app}\\
\alpha&\approx &  \mp\frac{1}{\rho_\lambda}\frac{1}{\gamma K}\left(\frac{P}{K}\right)^{(2-\gamma)/\gamma} \\
\beta &\approx &  \frac{6r^2[r-2M(r)]}{(1-\alpha)^2}\frac{1}{(\gamma K \rho_\lambda)^2}\left(\frac{P}{K}\right)^{(3-2\gamma)/\gamma}
\end{eqnarray}
With these results it is easy to check that, when $P\to 0$ and for $\gamma<2$, $\alpha \to 0$ and $\beta P^{(0)}_r \to 0$, and (\ref{eq:newP}) boils down to $P_r\approx -P_r^{(0)}\propto \left(\frac{P}{K}\right)^{1/\gamma}$, which is well-behaved and vanishes in this limit. Knowing the behavior of $P_r$, it is easy to verify that $f'_P P_r$, $P_{rr}$, and $f'_P P_{rr}$ are well-behaved for $\gamma<2$. However, the term $f'_{PP}P_r^2$ of (\ref{eq:psi-P}) and (\ref{eq:Mr-P}) is problematic \footnote{ The problems introduced by this term  via (\ref{eq:psi-P}) and (\ref{eq:Mr-P}) do not spoil the behavior of $P_{rr}$, which is well-behaved for $\gamma<2$.}. In fact, from (\ref{eq:f'PP_app}) and (\ref{eq:P0_app}) it follows that
\begin{equation}\label{eq:div}
f'_{PP}P_r^2\approx \mp\frac{2}{\rho_\lambda}\frac{(1-\gamma)}{(\gamma K)^2}\frac{M_{tot}^2}{r^2(r-2M_{tot})^2}\left(\frac{P}{K}\right)^{(3-2\gamma)/\gamma} \ . 
\end{equation}
This term diverges for $\gamma>3/2$  when $P\to 0$. The existence of this divergence was first reported in \cite{Sot08a}, where the matching conditions between the internal and the external (Schwarzschild) metric were analyzed. Since this term is contained in $\psi_r$ and $M_r$, and such terms appear in the definition of the Riemann tensor, it was concluded that the geometry becomes singular near the surface of polytropes with index $2<\gamma<3/2$. Since the physically interesting case $\gamma=5/3$ (degenerate, non-relativistic fermion gas) lies within this interval, this result was regarded as a serious theoretical concern about the viability of Palatini $f(R)$ theories. We will study now whether the (mathematical) divergence found in (\ref{eq:div}) is physical or not and how stringent is its effect on the viability of Palatini $f(R)$ theories.\\

First of all, for the models considered here, the divergence of $f'_{PP}P_r^2$ is entirely due to the contribution coming from the polytropic equation of state.  This can be seen from the definition of $f'_{PP}$ in (\ref{eq:f_PP}) [or from (\ref{eq:f'pp}) or (\ref{eq:f'PP_app})] since $d\rho/dP$ and $d^2\rho/dP^2$ diverge as $P\to 0$ irrespective of the function $f(R)$. If $d\rho/dP$ and $d^2\rho/dP^2$ were finite as $P\to 0$, the divergence would not arise. Second, a glance at the dimensionless (and exact) contribution $\frac{\rho_{PP}}{\rho_\lambda}(\rho+P)^2$ of $f'_{PP}P_r^2$ indicates that this term is strongly suppressed everywhere by the factor $1/\rho_\lambda\sim 10^{-92}$. (This also happens with all other correcting terms, which indicates that for our Planck scale corrected theory the interior structure of stars is virtually the same as in GR. In \cite{Sot08b}, however, the value of $\lambda$ was chosen in such a way that the interior of compact stars was indeed affected by the full $f(R)$ dynamics.) The exception occurs strictly in the limit $P\to 0$. Re-expressing that dimensionless factor in terms of the density, we find
\begin{equation}
\frac{\rho_{PP}}{\rho_\lambda}(\rho+P)^2\approx \frac{(1-\gamma)c^4}{\rho_\lambda(\gamma K )^2}\rho^{(3-2\gamma)}.
\end{equation}
Since for $\gamma>3/2$ this term diverges as $\rho\to 0$, it is worth looking at the value at which it becomes of order unity. This happens at $\rho=\rho_s$, where 
\begin{equation}\label{eq:rho_s}
\rho_s= \left(\frac{K^2\rho_\lambda}{c^4}\right)^{\frac{1}{3-2\gamma}} \ .
\end{equation}
For non-relativistic neutrons, $\rho\ll 6\cdot 10^{15} \ g/cm^3$, the ideal Fermi gas approximation leads to a polytropic equation of state of index $\gamma=5/3$ and $K=(3^{2/3}\pi^{4/3}/5)(\hbar^2/m_n^{8/3})\approx 5\cdot 10^9$ (in c.g.s. units), where $m_n$ is the neutron mass. The resulting $\rho_s$ is on the order of $\rho_s\sim 
10^{-210} \ g/cm^3$ \footnote{For non-relativistic electrons in white dwarfs, $\rho\ll 10^6 \ g/cm^3$, one finds $K\approx 10^13 Y_e^{5/3}$, where $Y_e$ is the mean number of electrons per baryon. This leads to $\rho_s\approx 5\cdot 10^{-196} Y_e^{-5} \ g/cm^3$.}. This density is, by far, well below any physical density one can imagine. In fact, for a free electron whose wave function is spread over the entire universe, the ratio $m_{e}/R^3_{Univ}$ is on the order of $\sim 10^{-118} \ g/cm^3$. This means that a single electron outside of this idealized polytrope is more than enough to cure this singularity (of course, we assume the electron wavefunction uniformly spread over a spherical shell to respect the symmetry of the problem) since its mere presence rises the average matter density $92$ orders of magnitude above the critical scale $\rho_s$.  Therefore, the existence of a curvature singularity at such extremely low densities should be regarded as unphysical, as an artifact of the idealized equation of state used. One should have in mind that an accurate EoS at laboratory densities (let alone at the density scale found here) is very complicated to derive, because electrostatic interactions and other subtle effects mask the simpler statistical effects of the idealized Fermi gas approximation \cite{Shapiro&Teukolsky}. The polytropic EoS should therefore be used within its expected regime of validity.

\section{Discussion and Conclusions}

This work was motivated by an extremely interesting and puzzling result: the existence of curvature singularities in regions of very low curvature  affecting all families of $f(R)$ lagrangians in Palatini formalism. If such singularities were physical, all $f(R)$ lagrangians except that of Hilbert-Einstein should be ruled out. And that would be so for arbitrarily small departures from $f(R)=R$. Such a result would have had very strong physical consequences, since it would have singled out GR as {\it the} privileged $f(R)$ lagrangian. Our analysis, however, shows that the (mathematical) low density/curvature singularities found in \cite{Sot08a} depend intimately on the particular EoS chosen and have different strength depending on the $f(R)$ model chosen. In the model that we studied, characterized by the Planck length squared, a single electron would suffice to cure all stellar singularities in the Universe. If instead of the Planck length we had chosen a larger length scale (smaller density scale $\rho_\lambda$), the surface singularity could occur at higher densities, perhaps within the regime of validity of the polytropic EoS,  and could not be removed by such simple means. In this sense,  the doubts raised in \cite{Sot08a} and \cite{Sot08b} about the viability of Palatini $f(R)$ models are well justified. In those works, the value of $\lambda$ was chosen on phenomenological grounds to be $\lambda=(0.15 \ km )^2$, which leads to $\rho_\lambda\sim 2\cdot 10^{18} \ g/cm^3$. This density scale is enough to pass all solar system weak field tests \cite{Olmo05} since $f'\approx 1$ to high accuracy everywhere. However, it leads to $\rho_s\approx 4\cdot 10^{12} \ g/cm^3$, which indicates that the divergent terms of $M_r$ and $\psi_r$ begin to grow well within the region of validity of the polytropic EoS. The study of polytropes carried out here and in \cite{Sot08a} and \cite{Sot08b} can thus be used to place bounds on the parameter $\lambda$ of the model under consideration and also in other models. If, for instance, we assume that the polytropic EoS should not be trusted below $\rho_s \sim 10^{-n} \ g/cm^3$, we find that 
\begin{equation}
\lambda\ll 10^{4-n/3} \ cm^2, 
\end{equation}
which places explicit constraints on the allowed values for $\lambda$. \\

To conclude, the important point was to show that not {\it all} lagrangians are ruled out by this theoretical experiment with polytropes. We have confirmed that the surface singularities exist, but their physical character depends on the details of the model and, as expected, ultraviolet corrections are allowed for suitable choices of parameters. In particular, Planck scale corrected models are save from such singularities. This is of great importance to the light of the recently discovered relation existing between non-perturbative approaches to quantum gravity and Palatini $f(R)$ theories \cite{Olmo-Singh08}. It turns out that the only consistent homogeneous and isotropic cosmology that can be constructed using the techniques of Loop Quantum Gravity and which is free of the Big Bang singularity \cite{Corichi-Singh} can be put into correspondence with a Palatini $f(R)$ lagrangian with ultraviolet, Planck scale, corrections. This provides a fundamental and so far non-existing theoretical justification for Palatini $f(R)$ theories, which have only been studied on phenomenological grounds in relation with the cosmic speedup. We hope that this new avenue of research helps us better understand the properties of quantum gravity and space-time near cosmological and black hole singularities using the more familiar techniques of modified gravities. \\

The author thanks E.Barausse, T.P. Sotiriou, and J.C. Miller for their critical reading of this manuscript and clarifications on their works. Special thanks go to F.Barbero for very useful comments and discussions on this subject. This work has been supported by MICINN  through a Juan de la Cierva postdoctoral contract.


\begin{thebibliography}{99}



\bibitem{WIL-liv05}
Clifford M. Will, {\it Living Rev.Rel.} 9,3,(2005), gr-qc/0510072 ; 

\bibitem{Tonry-Knop}
J. L. Tonry et al., {\it Astrophys. J.} {\bf 594}, 1 (2003); R. A. Knop
et al., {\it Astrophys. J.} {\bf 598}, 102 (2003).

\bibitem{Dark-energy}
T. Padmanabhan, {\it Phys.Rept.} 380 (2003) 235-320,hep-th/0212290; 
Peebles P.J.E. and B.Ratra {\it Rev.Mod.Phys.} 75, 559 (2003), astro-ph:0207347.

\bibitem{mod-grav-basic}
S.M.Carroll, V.Duvvuri, M.Trodden and M.S.Turner,{\it Phys.Rev.}{\bf D70} (2004) 043528, astro-ph/0306438; 
S. Capozziello, {\it Int. J. Mod. Phys.} D, 11, 483, 2002. 

\bibitem{mod-grav-long}
D.N. Vollick, {\it Phys. Rev.}{\bf D} 68, 063510 (2003);
S. Capozziello et al. {\it Int.J.Mod.Phys.}{\bf D12} (2003) 1969-1982;
T.Chiba, {\it Phys.Lett.}{\bf B}576 (2003) 5-11, astro-ph/0307338;
E.E.Flanagan, {\it Phys.Rev.Lett.}{\bf 92}, 071101 (2004);
D.N. Vollick, {\it Class.Quant.Grav.} {\bf 21}, 3813 (2004);
E.E.Flanagan, {\it Class.Quant.Grav.} {\bf 21}, 3817 (2004).
X. Meng, P. Wang, {\it Gen.Rel.Grav.} {\bf 36},1947,(2004);
G. Allemandi et al., {\it Gen. Rel. Grav.} {\bf 37},1891 (2005);
T.P. Sotiriou, {\it Gen.Rel.Grav} {\bf 38}, 1407 (2006), gr-qc/0507027;
Bao Li, M.-C. Chu, {\it Phys.Rev.} {\bf D} 74, 104010, (2006);
M. Amarzguioui et al., {\it Astron.Astrophys.} {\bf 454}, 707, (2006);
N.J. Poplawski, {\it Phys.Rev.} {\bf D} 74, 084032, (2006);
T.P. Sotiriou, {\it Class.Quant.Grav.} {\bf 23}, 1253,(2006);
T.P. Sotiriou, {\it Phys.Rev.} {\bf D}73,063515,(2006);
T. Koivisto, {\it Phys. Rev.} {\bf D} 73, 083517 (2006); 
T. Koivisto, {\it Class.Quant.Grav.} 23, 4289 (2006);
M.L. Ruggiero and L.Orio, {\it JCAP} 0701 (2007) 010, gr-qc/0607093; 
G.J. Olmo, {\it Phys. Rev.} D {\bf 75}, 023511 (2007);
B. Li, J.D.Barrow, and D.F. Mota, {\it Phys.Rev.} D {\bf 76},104047,(2007);
S.Fay, R. Tavakol and S. Tsujikawa, {\it Phys.Rev.} {\bf D} 75, 063509 (2007);
T.P. Sotiriou, {\it Phys.Lett.}B 664,225-228,(2008);
D.Saez-Gomez, arXiv:0809.1311 ;
B.Li, D.F.Mota, and D.J. Shaw, arXiv:0805.3428; 
V.Faraoni, arXiv:0810.2602.

\bibitem{f_reviews}

T.P.Sotiriou and V.Faraoni, arXiv:0805.1726; 
S.Nojiri and S.Odintsov, {\it Int. J. Geom. Meth. Mod. Phys.} 4, 115-146, (2007).

\bibitem{Sot08a} 
E. Barausse, T.P. Sotiriou, and J.C. Miller, {\it Class.Quant.Grav.} 25,062001(2008).

\bibitem{Kainu07a}
K.Kainulainen, V. Reijonen, and D. Sunhede, {\it Phys.Rev.} D {\bf 76},043503(2007)

\bibitem{Kainu07b}
K.Kainulainen, J.Piilonen, V. Reijonen, and D. Sunhede, {\it Phys.Rev.} D {\bf 76},024020(2007)

\bibitem{Sot08b} 
E. Barausse, T.P. Sotiriou, and J.C. Miller, {\it Class.Quant.Grav.} 25,105008(2008). 

\bibitem{Olmo07}
G.J. Olmo, {\it Phys. Rev. Lett.} {\bf 98}, 061101 (2007).

\bibitem{Olmo08a}
G.J. Olmo, {\it Phys. Rev.} D {\bf 77}, 084021 (2008).


\bibitem{Shapiro&Teukolsky}
S.L.Shapiro and S.A.Teukolsky, ``Black Holes, White Dwarfs, and Neutron Stars: the Physics of Compact Stars'', Wiley \& Sons (1983). 

\bibitem{Olmo05}
G.J. Olmo, {\it Phys. Rev. Lett.} {\bf 95}, 261102 (2005); 
G.J. Olmo, {\it Phys. Rev.} D {\bf 72}, 083505 (2005).

\bibitem{Olmo-Singh08}
G.J. Olmo and P.Singh, ``Effective Action for Loop Quantum Cosmology a la Palatini'', arXiv:0806.2783.

\bibitem{Corichi-Singh}
A.Corichi and P.Singh, `` Is loop quantization in cosmology unique?'', arXiv:0805.0136 .

\end{thebibliography}
\end{document}